\documentclass[a4paper,twoside]{article}
\usepackage[dvips]{graphicx}
\usepackage{latexsym,epsf}
\usepackage{multicol}
\usepackage[centertags]{amsmath}
\usepackage [cp1251]{inputenc}
\usepackage [ukrainian, russian, english]{babel}
\usepackage {fancyhdr}

\renewcommand {\footrule}{\vbox to 0pt{\hbox to \headwidth{ \hrulefill \hspace{63mm}}\vss}}

\makeatletter
\renewcommand{\ps@plain}{
\renewcommand{\@oddhead}{}
\renewcommand{\@evenhead}{}
\renewcommand{\@oddfoot}{\hfil \thepage}
\renewcommand{\@evenfoot}{\thepage \hfil \hfil}}
\makeatother 

\makeatletter
\renewcommand{\@biblabel}[1]{#1.\hfill}
\makeatother
\title{\textbf{\Large POLARIZED BREMSSTRAHLUNG IN THE EQUIVALENT PHOTON METHOD}}
\author{\textbf{\textit{M.V. Bondarenco\footnote{\normalfont
Corresponding author E-mail address: bon@kipt.kharkov.ua} }}
\\
\emph{\small National Science Center "Kharkov Institute of
Physics and Technology" 61108 Kharkov, Ukraine}\\
{\small(Received April 21, 2008)}}
\begin{document}
\selectlanguage{english}
\date{}
\maketitle

\thispagestyle{fancy}

\begin{center}
\begin{minipage}{165mm}
{\small
An economic technique for calculation of polarized bremsstrahlung
process is proposed, assuming typical atomic momentum transfer $q\ll
m$. The adopted approach is based on the natural reduction of the
matrix element to the form $V^{\alpha}\gamma^{\alpha}+A_{5}\gamma
^{5}$. Polarization distribution in the fully differential
cross-section is analyzed. It is found that at a given momentum
transfer to the atom polarization in the plane of small radiation
angles is oriented along circles passing through two common points.
It is shown that with angular selection of radiated photons carried
out, even with momentum transfers to the atom being integrated over,
for particular radiation angles polarization may stay as high as
100\%. Without angular selection of photons, only by control of the
recoil, it is impossible to gain radiation polarization above 50\%.}
\par \vspace{1ex}
\end{minipage}
\end{center}
\begin{center}
\textbf{\textsc{1. INTRODUCTION}}
\end{center}
Radiation from ultra-relativistic electrons is naturally collimated
along the forward direction, with the opening angle being inversely
proportional to the electron energy, but up to energies $\sim10$
GeV, corresponding to typical radiation angles $\sim 10^{-4}$ rad,
angular distribution of radiation can be resolved. If that is done
in practice, photon polarization effects necessarily come into play.
In particular, the connection between the polarization degree and
the angular distribution is important for purposes of preparation of
polarized photon beams. The issue of polarization account may arise
also at investigation of spatial evolution of electromagnetic
showers. Calculations of differential cross-section for the
bremsstrahlung process were conducted in various frameworks
\cite{Ter-Mik}, \cite{BKF}, but acquired reputation of rather
cumbersome a subject. Presentation of the final results in the
literature does not help gaining detailed intuition.

The conditions of radiation from ultra-relativistic electrons in
matter are such that small momentum transfers to atoms dominate
($q\sim r_{B}^{-1}\ll m$). Under such conditions the expression for
bremsstrahlung cross-section, called dipole approximation, is
similar to Compton scattering cross-section from the standpoint of
the equivalent photon method. As a matter of fact, the latter method
is usually used to derive characteristics of bremsstrahlung,
averaged over directions of emission and over polarizations of
final, and also (through azimuthal integration in momentum
transfers) over polarizations of initial quanta. Nevertheless, it
can be extended to the polarized case, provided one traces
correspondence of polarizations (that is traditionally achieved via
a transition between reference frames \cite{Ter-Mik}).

From the technical side, the case of Compton scattering has the
advantage that photon polarization vectors in it might be chosen
orthogonal simultaneously to two momenta in the problem (out of
three momenta not bound by 4-momentum conservation). That is
expected to substantially simplify the procedure of calculation with
the account of polarizations and, in the conventional technology of
cross-section calculation via computation of a spur from the squared
matrix amplitude \cite{Feyn}, this is indeed the case. Despite the
simplifications gained, calculations for this process still are of
formidable complexity. This is in mark contrast with the concise
final result, and should be blamed on nothing but poor efficiency of
the method of straightforward spurring.

As was discussed in \cite{VANT}, a more adequate method of
calculation of QED-processes, also offering access to fermion
polarization observables, is the evaluation of spin amplitudes in a
matrix basis chosen by some convenience reasons.

In the present paper, firstly, it is desired to apply the spin
amplitude approach to the specific case of Compton scattering. It
appears that in this particular case the most convenient spin matrix
choice is different from what may be appropriate in other cases.
With the choice adopted in the present article, the whole procedure
of differential cross-section calculation with the account of
initial and final photon polarizations becomes fairly elementary.
Issuing from the representation for the differential cross-section
for Compton scattering in a gauge- and Lorenz-invariant form, it is
straightforward further on to pass to the differential cross-section
of bremsstrahlung in laboratory frame, corresponding to conditions
of peripheral scattering, i.e. dipole approximation. This provides
an alternative to explicit implementation of the equivalent photon
method.

The second objective of the present work is to discuss the features
of polarization distribution as a function of the radiation angle at
fixed transverse direction of momentum transfer to the atom. The
characteristic feature is that the polarization is aligned along
circles, passing through two knot points in the space of radiation
angles. One of the circles has its centre coinciding with the origin
- and that has an important consequence: upon averaging over
momentum transfers to the atom at radiation direction fixed, only
contributions with identical direction of polarization are summed
up. And since at small radiation frequencies the polarization in the
doubly differential cross-section is close to 100\%, after the
averaging it remains near so.

\vspace{3mm}
\begin{center}
\textbf{\textsc{2. CALCULATION OF AMPLITUDES AND CROSS-SECTION FOR
COMPTON SCATTERING}}
\end{center}
We are considering the process of electron scattering on an
atom\footnote{A composite and lower-dimensional scattering system
will be addressed in Sec. 3.1.}, accompanied by emission of a single
photon. Denoting by $p$ and $p'$ 4-momenta of the initial and the
final electrons, $q$ and $q'$ - the momentum transfer to the atom
and the momentum of the emitted photon, the 4-momentum conservation
and mass shell conditions for them read as:
\begin{equation}
   p+q=p'+q', \quad p^{2}=p'^{2}=m^{2}, \quad q'^{2}=0.\nonumber
\end{equation}
For $q^{2}$ there is no strict condition, but
\begin{equation}\label{cond}
   q^{2}\sim r_{B}^{-2}\sim e^{4}m^{2}\ll m^{2},
\end{equation}
$r_{B}$ being Bohr radius. This estimate relates mainly to $q$
components orthogonal to $p$ . The component $q_{z}$ parallel to $p$
is small, as long as typical denominators emerging in Feynman
diagrams are of order $pq\simeq Eq_{z}\sim m^{2}$, and $E$ is large.
Then $q_{z}r_{B}\ll 1$, owing to which condition the matrix element
for the whole process factorizes into Born-level radiation matrix
element $M_{rad}(q_{\perp},q')$ and the exact elastic scattering
amplitude $A_{scat}(q_{\perp})$:
\begin{eqnarray}\label{Mrad}
  &\,& T_{fi}=A_{scat}\left(q_{\perp}\right)\sqrt{4\pi}e M_{rad}\left(q_{\perp},q'\right),\nonumber\\
  &\,& d\sigma_{scat}=\left|A_{scat}\right|^{2}\frac{ d^{2}q_{\perp} }{(2\pi)^{2}},\nonumber \\
  &\,& M_{rad}\!=\!\bar{u}_{p'}\!\!\! \left(\! \frac{\hat{e}'^{*}\!(\widehat{p}+\!\! \widehat{q}\!+\!m)\widehat{e}}{2pq}\! - \! \frac{\widehat{e}(\widehat{p}-\! \widehat{q}'\!+\!m)\hat{e}'^{*}}{2pq'} \!\right)\!\! u_{p},\\
  &\,& d\sigma_{rad}=\frac{1}{2E}\left|T_{fi}\right|^{2}\frac{d^{2}q_{\perp}}{(2\pi)^{2} 2E'}\frac{d^{3}q'}{(2\pi)^{3} 2q'_{0}}\nonumber
\end{eqnarray}\\
(the elastic scattering amplitude through small angles can be
regarded as independent from the electron polarization). We are
interested in $M_{rad}(q_{\perp},q')$ modulus squared and averaged
over initial electron and summed over final electron polarizations.
By the straightforward Feynman's method, this quantity is to be
computed as
\begin{equation}\label{Spur}
\left\langle \left|M_{rad} \right|^{2} \right\rangle
\!=\!\frac{1}{2}Sp\left(\widehat{p}'\!+\! m\right)\!\!\left(
\frac{\hat{e}'^{*}(\widehat{p}+\widehat{q}+m)\widehat{e}}{2pq} \!-\!
\frac{\widehat{e}(\widehat{p}-\widehat{q}'+m)\hat{e}'^{*}}{2pq'}
\right)\!\!\left(\widehat{p}+\!m\right)\!\!\left(
\frac{\hat{e}^{*}(\widehat{p}+\widehat{q}+m)\widehat{e}'}{2pq} \!-\!
\frac{\widehat{e}'(\widehat{p}-\widehat{q}'+m)\hat{e}^{*}}{2pq'}
\right)\!\!.
\end{equation}
\noindent However, at that one needs to calculate a spur from a
polynomial of 8th degree in $\gamma$-matrices. Typically, even
relying on properties of photon crossing symmetry and advantages of
the gauge choice (see (\ref{gauge}) below), that entails
calculations along 2-3 pages (cf. \cite{Feyn}). A more efficient
approach would be prior reduction of the spin matrix to some
"minimal" form, embarking on the condition of initial and final
electron bispinors belonging to the mass shell.

\textbf{\textsc{2.1. Deduction of basic amplitudes.}}

As usual, the computations of Compton scattering are strongly
facilitated in the gauge
\begin{equation}\label{gauge}
ep=eq=0=e'p=e'q'.
\end{equation}
To start with, commute in the matrix element (\ref{Mrad})
$\widehat{p}$ through $\widehat{e}$ , $\widehat{e}'$ to a position
neighboring to $u_{p}$:
\begin{eqnarray}\label{init}
M_{rad}\!\!\!&=&\!\!\!\bar{u}_{p'}\!\! \left(
\frac{\hat{e}'^{*}(\widehat{p}+\widehat{q}+m)\widehat{e}}{2pq} -
\frac{\widehat{e}(\widehat{p}-\widehat{q}'+m)\hat{e}'^{*}}{2pq'}
\right)\!\! u_{p} \nonumber \\
\!\!\!&\cong & \!\!\!\bar{u}_{p'}\!\! \left(
\frac{\hat{e}'^{*}\widehat{q}\widehat{e}}{2pq} +
\frac{\widehat{e}\widehat{q}'\hat{e}'^{*}}{2pq'} \right)\!\! u_{p}.
\end{eqnarray}
Feynman in \cite{Feyn} started squaring from this modified
representation, but it is still 8-th order in $\gamma$-matrices, and
we shall proceed a little further. With the application of the
standard formula
\begin{equation*}
\gamma^{\alpha}\gamma^{\beta}\gamma^{\gamma}=g^{\alpha\beta}\gamma^{\gamma}-g^{\alpha\gamma}\gamma^{\beta}+g^{\beta\gamma}\gamma^{\alpha}+i\varepsilon^{\nu\alpha\beta\gamma}\gamma^{\nu}\gamma^{5},
\end{equation*}
(\ref{init}) naturally reduces to a basic-matrix form:
\begin{equation}\label{red}
M_{rad}=\bar{u}_{p'}
\left(V^{\alpha}\gamma^{\alpha}+A_{5}\gamma^{5}\right)u_{p}.
\end{equation}
The coefficients in the decomposition are
\begin{eqnarray*}
&\,& V^{\alpha}=\frac{e^{\alpha}qe'^{*}-q^{\alpha}ee'^{*}}{2pq}+\frac{e'^{\alpha*}q'e-q'^{\alpha}ee'^{*}}{2pq'},\\
&\,& A_{5}=-i\varepsilon^{\mu\alpha\beta\gamma}\frac{p^{\mu}}{m}e^{\alpha}e'^{\beta *}G^{\gamma}, \\
&\,& \left(
G^{\gamma}=\frac{q^{\gamma}}{2pq}-\frac{q'^{\gamma}}{2pq'}\right)
\end{eqnarray*}
and it was utilized that vector $G$, as well as $e$, $e'$, is
orthogonal to $p$,
\begin{equation*}
Gp=0,
\end{equation*}
so,
\begin{equation*}
\varepsilon^{\nu\alpha\beta\gamma}e^{\alpha}e'^{\beta*}G^{\gamma}=\frac{p^{\nu}}{m}\cdot\varepsilon^{\mu\alpha\beta\gamma}\frac{p^{\mu}}{m}e^{\alpha}e'^{\beta*}G^{\gamma},
\end{equation*}
whereas action of $p^{\nu}\gamma^{\nu}\gamma^{5}/m$ on $u_{p}$ gives
$-\gamma^{5}$.

One can still add to the vector $V^{\alpha}$  an arbitrary vector,
proportional to $(p-p')^{\alpha}$. It is advantageous to tune it so
that $V$ be orthogonal to momentum $p$:
\begin{eqnarray*}
V^{\alpha}\rightarrow
V_{p}^{\alpha}&=&\frac{e^{\alpha}qe'^{*}-q^{\alpha}ee'^{*}}{2pq}+(p-p')^{\alpha}\frac{ee'^{*}}{(p-p')^{2}}
+\frac{e'^{\alpha*}q'e-q'^{\alpha}ee'^{*}}{2pq'}+(p-p')^{\alpha}\frac{ee'^{*}}{(p-p')^{2}} \\
&=&e^{\alpha}e'^{*}G-e'^{\alpha*}eG-G^{\alpha}ee'^{*}\frac{pq+pq'}{m^{2}-pp'}.
\end{eqnarray*}
Now $V_{p}^{\alpha}$ and $A_{5}$ together have 4 independent
components, as it should be for parametrization of a matrix
describing transition between two spin-1/2 on-shell states.

\vspace{3mm}
 \textbf{\textsc{2.2. Computation of the differential
cross-section, averaged over fermion polarizations.}}

Substituting (\ref{red}) to (\ref{Spur}), the spur is calculated
easily:
\begin{equation*}
\left\langle \left| M_{rad} \right| ^{2} \right\rangle
=\frac{1}{2}Sp\left(\widehat{p}'+m\right)\left(
V_{p}^{\alpha}\gamma^{\alpha}+A_{5}\gamma^{5}
\right)\left(\widehat{p}+m\right)\left(
V_{p}^{\alpha*}\gamma^{\alpha}+A_{5}^{*}\gamma^{5} \right)
=2(m^{2}-pp')\left\{ \left|V_{p}\right|^{2}-\left|A_{5}\right|^{2}
\right\}.
\end{equation*}
The entries thereat are evaluated to be
\begin{equation*}
\left|V_{p}\right|^{2}=\left| e^{\alpha}e'^{*}G-e'^{\alpha*}eG
\right|^{2}+G^{2}\left|ee'^{*}\right|^{2}\left(\frac{pq+pq'}{m^{2}-pp'}\right)^{2},
\end{equation*}
\begin{eqnarray*}
-\left|A_{5}\right|^{2}=\left| %
\begin{array}{ccc}
  \left|e\right|^{2} & ee' & eG \\
  (ee')^{*} & \left|e'\right|^{2} & e'^{*}G \\
  e^{*}G & e'G & G^{2} \\
\end{array}%
\right|\!\!&=&\!\!\left|e\right|^{2}\left|e'\right|^{2}G^{2}+2
\mathcal{R}e \, ee' \cdot e'^{*}G\cdot
e^{*}G-\left|e\right|^{2}\left|e'G\right|^{2}-\left|e'\right|^{2}\left|eG\right|^{2}-G^{2}\left|ee'\right|^{2}\\
\!\!&=&\!\!-\left| e^{\alpha}e'^{*}G-e'^{\alpha*}eG
\right|^{2}+\left|e\right|^{2}\left|e'\right|^{2}G^{2}-G^{2}\left|ee'\right|^{2}.
\end{eqnarray*}
In sum, after cancellation of terms $\pm \left|
e^{\alpha}e'^{*}G-e'^{\alpha*}eG \right|^{2}$,
\begin{equation*}
\left|V_{p}\right|^{2}-\left|A_{5}\right|^{2}=\left|e\right|^{2}\left|e'\right|^{2}G^{2}-G^{2}\left|ee'\right|^{2}+G^{2}\left|ee'^{*}\right|^{2}\left(\frac{pq+pq'}{m^{2}-pp'}\right)^{2},
\qquad G^{2}=\frac{m^{2}-pp'}{2pq\cdot pq'},
\end{equation*}
\begin{equation}\label{KN1}
\left\langle \left| M_{rad} \right| ^{2} \right\rangle
=\frac{1}{pq\cdot pq'}\left\{ \left(
\left|e_{p}\right|^{2}\left|e'_{p}\right|^{2}-\left|e_{p}e'_{p}\right|^{2}\right)(pq-pq')^{2}+\left|e_{p}
e'^{*}_{p}\right|^{2}(pq+pq')^{2} \right\}.
\end{equation}
(In the final formula an explicit subscript $p$ at polarization
vectors is introduced emphasizing the employed gauge).

In what follows, we will be mainly interested in the case of
linearly polarized initial photons. Then, the final photon
polarization is also linear. For those conditions one can set
$\left|e_{p} e'^{*}_{p} \right|=\left|e_{p}e'_{p}\right| =\left(
e_{p}e'_{p} \right)$ and add in (\ref{KN1}) the two like terms:
\begin{equation}\label{KN2}
\left\langle \left| M_{rad} \right| ^{2} \right\rangle =e_{p}^{2}
e'^{2}_{p} \frac{(pq-pq')^{2}}{pq\cdot pq'}+4 (e_{p} e'_{p})^{2}.
\end{equation}
This is the renowned Klein-Nishina's formula for linearly polarized
initial and final photons \cite{Feyn}, \cite{BLP}.

To apply formula (\ref{KN2}) to bremsstrahlung in laboratory frame,
where the scatterer atom is at rest, it should first be rendered a
gauge-invariant appearance. To this end, substitute for $e_{p}$,
$e'_{p}$ expressions $e_{p}=e-q\frac{ep}{pq}$,
$e'_{p}=e'-q'\frac{e'p}{pq'}$, where $e$, $e'$ are polarization
vectors in an arbitrary gauge:
\begin{equation}\label{KN3}
\left\langle \left| M_{rad} \right| ^{2} \right\rangle
=\left(e-q\frac{ep}{pq}\right)^{2}\left(e'-q'\frac{e'p}{pq'}\right)^{2}\frac{(pq-pq')^{2}}{pq\cdot
pq'}+4\left\{\left(e-q\frac{ep}{pq}\right)\left(e'-q'\frac{e'p}{pq'}\right)\right\}^{2}.
\end{equation}
Thereupon, this formula can be applied in the laboratory frame.

\begin{center}
\vspace{3mm}
\textbf{\textsc{3. APPLICATION TO THE BREMSSTRAHLUNG IN
LABORATORY FRAME.}} \end{center} For bremsstrahlung in the
laboratory frame $e=(1,\textbf{0})$, $e'=(0,\bf{e'})$,
$q=(0,-\bf{q})$, $q'=(\omega,\bf{k})$. At that, combinations
$e_{p}$, $e'_{p}$ in components equal
\begin{equation}\label{comp1}
e-q\frac{ep}{pq}\simeq \left(
1,1,\frac{\bf{q}_{\perp}}{q_{z}}\right),
\end{equation}
\begin{equation}\label{comp2}
e'\!-q'\frac{e'p}{pq'}\simeq \!\!\left(
e'_{z}\omega\frac{E}{E'q_{z}},e'_{z}\!\!\left(1\!+\!\omega\frac{E}{E'q_{z}}\right)\!\!,\bf{e}'_{\perp}\!+\!\bf{k}_{\perp}\frac{\bf{e}'\bf{p}}{E'q_{z}}\right)\!\!,
\end{equation}
In terms of orthogonal components, the combination $\bf{e}'\bf{p}$
entering (\ref{comp2}) equals $\textbf{e}'\textbf{p}\simeq
\frac{E}{\omega}(\textbf{e}'\textbf{k}-\textbf{e}'_{\perp}\textbf{k}_{\perp})=-\frac{E}{\omega}\textbf{e}'_{\perp}\textbf{k}_{\perp}$.

Apparently, in scalar products present in (\ref{KN3}), the temporal
and the longitudinal spatial components of vectors
(\ref{comp1}-\ref{comp2}) do not essentially contribute, except in
$e'^{2}_{p}$, which is easier calculated in the Lorenz-invariant
fashion, with the use of $e'q'=0$, $q'^{2}=0$:
$e'^{2}_{p}=e'^{2}=-1$. So,
\begin{eqnarray*}
e_{p}^{2}&\simeq &-\frac{\textbf{q}_{\perp}^{2}}{q_{z}^{2}},\\
e'^{2}_{p}&=& -1,\\
e_{p} e'_{p}&\simeq &
\frac{\left|\textbf{q}_{\perp}\right|}{q_{z}}\left(
-\textbf{n}_{\textbf{q}_{\perp}}+\frac{\left(\textbf{k}_{\perp}\textbf{n}_{\textbf{q}_{\perp}}\right)}{E'q_{z}}\frac{E}{\omega}\textbf{k}_{\perp}\right)\textbf{e}'
\end{eqnarray*}
with
\begin{equation*}
\textbf{n}_{\textbf{q}_{\perp}}=\frac{\textbf{q}_{\perp}}{\left|\textbf{q}_{\perp}\right|}.
\end{equation*}
Next, the kinematical combinations required are
\begin{equation*}
\frac{(pq-pq')^{2}}{pq\cdot pq'}\simeq \frac{(qq')^{2}}{pq\cdot
pq'}\simeq \frac{(\omega q_{z}')^{2}}{Eq_{z}\cdot
E'q_{z}}=\frac{\omega^{2}}{EE'},
\end{equation*}
\begin{eqnarray*}
&\,& E'q_{z}\simeq p'q\simeq pq'=E\omega-p_{z}k_{z}\\
&\,& \simeq E\omega-\!\!\left(
E-\frac{m^{2}}{2E}\right)\!\!\left(\omega-\frac{\textbf{k}_{\perp}^{2}}{2\omega}\right)
\!\simeq \!
\frac{\omega}{2E}m^{2}\!+\!\frac{E}{2\omega}\textbf{k}_{\perp}^{2},
\end{eqnarray*}
or, introducing the radiation angle
$\bf{\theta}_{\bf{k}}=\textbf{k}_{\perp}/\omega$ and ratios
 $\gamma=E/m\gg 1$, $x_{\omega}=\omega/E$, $0\leq x_{\omega}\leq 1$,
\begin{equation*}
q_{z}=\frac{mx_{\omega}}{2(1-x_{\omega})}\left(
\frac{1}{\gamma}+\gamma\bf{\theta}_{\textbf{k}}^{2}\right).
\end{equation*}

Substituting all the ingredients into (\ref{KN3}), one arrives at
the expression
\begin{equation}\label{F}
\left\langle \left| M_{rad} \right| ^{2} \right\rangle
\!=\!4\textbf{q}_{\perp}^{2} \frac{(1-x_{\omega})}{m^{2}\left(
\frac{1}{\gamma}+\!\gamma
\bf{\theta}_{\textbf{k}}^{2}\right)^{2}}\!\!\left\{\!1+\!\!\frac{4(1-x_{\omega})}{x_{\omega}^{2}}\left|\bf{\nu}\textbf{e}'
\right|^{2}\!\!\right\}.
\end{equation}
Here $\bf{\nu}$ is a vector
\begin{equation*}
\bf{\nu} = -\textbf{n}_{\textbf{q}_{\perp}}+\frac{ 2
}{\gamma^{-2}+\bf{\theta}_{\textbf{k}}^{2}}(\bf{\theta}_{\textbf{k}}\textbf{n}_{\textbf{q}_{\perp}})\bf{\theta}_{\textbf{k}},
\end{equation*}
along which the radiation polarization orients itself \footnote{It
may be worth pointing out, that except overall proportionality to
$\textbf{q}_{\perp}^{2}$, $\left\langle \left| M_{rad} \right| ^{2}
\right\rangle$ is also dependent on
$\textbf{n}_{\textbf{q}_{\perp}}$ despite $\textbf{q}_{\perp}^{2}\ll
m^{2}$ - unless it is integrated over directions of
$\bf{\theta}_{\textbf{k}}$ and summed over $\textbf{e}'$. That
circumstance is often missed in presentations of the equivalent
photon method, including treatises \cite{BLP}, \cite{AB}, though is
taken care in \cite{Ter-Mik}.}. The absolute value of the
polarization amounts
\begin{equation*}
P=\frac{\bf{\nu}^{2}}{\frac{x_{\omega}^{2}}{2(1-x_{\omega})}+\bf{\nu}^{2}},
\end{equation*}
with
\begin{eqnarray*}
\bf{\nu}^{2}&=&1-\frac{4\gamma^{-2}}{(\gamma^{-2}+\bf{\theta}_{\textbf{k}}^{2})^{2}}(\bf{\theta}_{\textbf{k}}
\textbf{n}_{\textbf{q}_{\perp}})^{2}\\
&\equiv&\frac{(\gamma \bf{\theta}_{\textbf{k}}
+\textbf{n}_{\textbf{q}_{\perp}})^{2}(\gamma
\bf{\theta}_{\textbf{k}}
-\textbf{n}_{\textbf{q}_{\perp}})^{2}}{(1+\gamma^{2}\bf{\theta}_{\textbf{k}}^{2})^{2}}.
\end{eqnarray*}

It is easy to show by solving the differential equation $d\theta
_{y}/d\theta_{x}=\nu_{y}\left(\theta
_{x},\theta_{y}\right)/\nu_{x}\left(\theta _{x},\theta_{y}\right)$,
that the curves, in every point tangential to the direction of
vector $\bf{\nu}$, are circles passing through two specific points:
$\bf{\theta}_{\textbf{k}}=\pm
\textbf{n}_{\textbf{q}_{\perp}}/\gamma$ (see Fig. 1). In those two
points the polarization turns to zero. In vicinities of those points
$\bf{\nu}^{2}\simeq (\gamma \bf{\theta}_{\textbf{k}}\mp
\textbf{n}_{\textbf{q}_{\perp}})^{2}$ . At distances from them much
greater then $\frac{x_{\omega}}{2\gamma\sqrt{1-x_{\omega}}}$
polarization is close to 100\%.
\\
\begin{minipage}{70mm}
\begin{center}
\includegraphics[width=\textwidth]{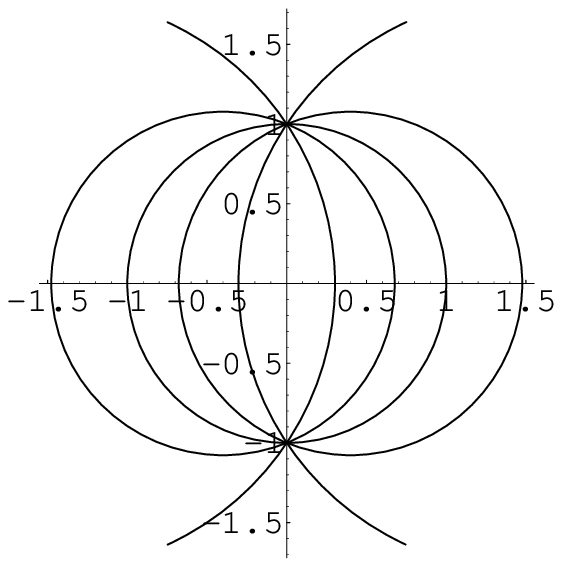}
\emph{\textbf{Fig.1.}} {\it Curves in $\gamma
\bf{\theta}_{\textbf{k}}$ plane, directing polarization of
radiation. All displayed curves are circles.
$\textbf{n}_{\textbf{q}_{\perp}}$ is a unit vector along the
vertical axis.}\\
\end{center}
\end{minipage}
\vspace{3mm}

\noindent The doubly differential cross-section in itself is usually
not measured in experiment, as long as recoiling atoms are not
detected. So, the picture described above serves mainly for
intuition purposes. Below we shall discuss two most important cases
of $A_{scat}$ dependence on $\textbf{n}_{\textbf{q}_{\perp}}$. In
both those cases there is a factorization of azimuthal and radial
dependences
$A_{scat}(\textbf{q}_{\perp})=\Phi(\textbf{n}_{\textbf{q}_{\perp}})A(\left|\textbf{q}_{\perp}
\right|)$, the latter being irrelevant for polarization effects in
view of homogeneity of $M_{rad}$ dependence on
$\textbf{q}_{\perp}^{2}$. Hence, it suffices to discuss averaging of
$\left\langle \left| M_{rad} \right| ^{2} \right\rangle$ over
$\textbf{n}_{\textbf{q}_{\perp}}$.
\vspace{3mm}

\textbf{\textsc{3.1. Planar geometry.}}

In the planar geometry $A_{scat}(\textbf{q}_{\perp})$ exhibits a
sharp peak along some particular direction
$\textbf{n}_{\textbf{q}_{\perp}}$.\footnote{Physically, it may
correspond either to electron passage through a thin crystal close
to a strong crystalline plane, or to passage through gap of a magnet
deflecting electrons to small angles. It should be minded that in
those cases actual dimensions of the scattering system exceed the
cross-section of electron beam in both transverse directions, so the
concept of differential cross-section looses direct physical sense.
Nevertheless, values of polarization extracted from it do not depend
on $A_{scat}$ and are correct.} If one selects photons away from
knot directions $\pm \textbf{n}_{\textbf{q}_{\perp}}/\gamma$, where
polarization is maximal - say, emitted in the middle band of angles
$\left|\bf{\theta}_{\textbf{k}}
\textbf{n}_{\textbf{q}_{\perp}}\right|<1/2\gamma$ (ref. to Fig. 1),
the dependence of polarization on $x_{\omega}$ may be estimated from
formula (\ref{F}) upon substitution in it $\textbf{k}_{\perp}\perp
\textbf{q}_{\perp}$:
\begin{eqnarray*}
\left\langle \left| M_{rad} \right| ^{2} \right\rangle
\!=\!\frac{4\textbf{q}_{\perp}^{2}(1-x_{\omega})}{m^{2}\!\left(\frac{1}{\gamma}+\gamma
\bf{\theta}_{\textbf{k}}^{2}\right)^{2}}
\left\{\!1\!+\!\frac{4(1-x_{\omega})}{x_{\omega}^{2}}\!\left|\textbf{n}_{\textbf{q}_{\perp}}\textbf{e}'
\right|^{2} \right\}, \,\,\,\,\,\,(\textbf{k}_{\perp}\perp
\textbf{q}_{\perp}).
\end{eqnarray*}
Then,
\begin{equation*}
P\simeq P(x_{\omega})=\frac{1}{1+x_{\omega}^{2}/2(1-x_{\omega})},
\quad (\left|\bf{\theta}_{\textbf{k}}
\textbf{n}_{\textbf{q}_{\perp}}\right|\ll 1/\gamma)
\end{equation*}
independently on $\bf{\theta}_{\textbf{k}}\times
\textbf{n}_{\textbf{q}_{\perp}}$. As Fig. 2 displays, polarization
stays higher then 90\% for $x_{\omega}<0.35$.
\\
\begin{minipage}{80mm}
\includegraphics[width=\textwidth]{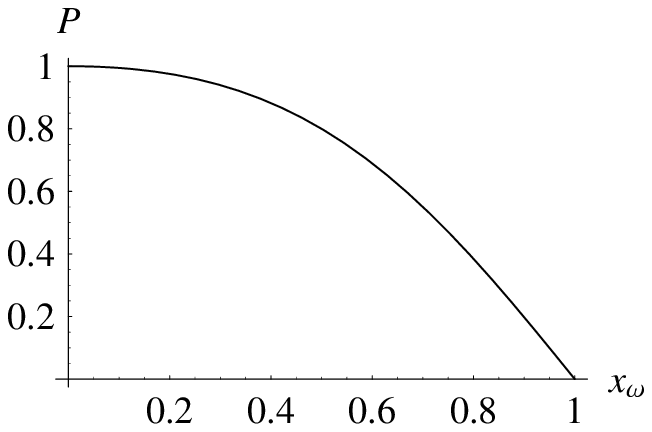}
\begin{center}
\emph{\textbf{Fig.2.}} {\it Polarization of bremsstrahlung at planar
scattering and orthogonal radiation $\left|
\bf{\theta}_{\textbf{k}}\textbf{n}_{\textbf{q}_{\perp}} \right|\ll
1/\gamma$.}\\
\end{center}
\end{minipage}
\vspace{3mm}

On the other hand, if angular separation is not attempted (which
might be technically challenging at $\gamma>10^{4}$), and only the
natural collimation due to emission from an ultra-relativistic
particle is used, one needs to integrate over radiation angles, or
equivalently, photon transverse momenta. Evaluation of the integral
of (\ref{F}) over $d^{2}\textbf{k}_{\perp}$ yields:
\begin{eqnarray*}
\int \left\langle \left| M_{rad} \right| ^{2} \right\rangle
\frac{d^{2}k_{\perp}}{(2\pi)^{2}} = \int \left\langle \left| M_{rad}
\right| ^{2} \right\rangle \frac{\omega^{2}d^{2}\theta _{\textbf{k}}
} {(2\pi)^{2}}
=\textbf{q}_{\perp}^{2}\gamma^{2}\frac{1-x_{\omega}}{\pi}
\left\{x_{\omega}^{2}+\frac{2}{3}(1-x_{\omega})\!\left[1+2(\textbf{n}_{\textbf{q}_{\perp}}\textbf{e}')^{2}\right]\!
\right\},
\end{eqnarray*}
\begin{equation*}
P\simeq
P(x_{\omega})=\frac{1-x_{\omega}}{\frac{3}{2}x_{\omega}^{2}+2(1-x_{\omega})}.
\end{equation*}
\\
\begin{minipage}{80mm}
\includegraphics[width=\textwidth]{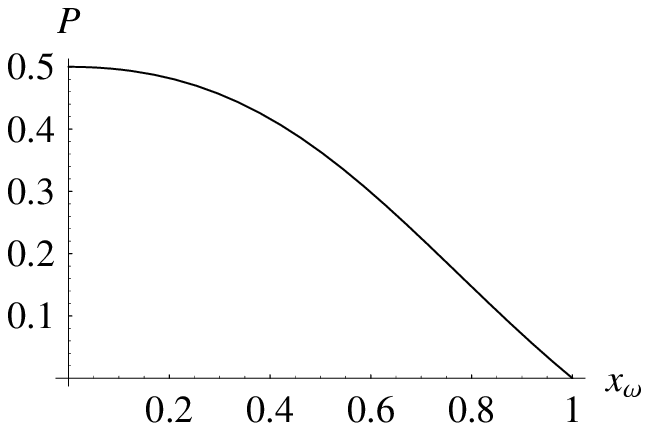}
\begin{center}
\emph{\textbf{Fig.3.}} {\it Polarization of bremsstrahlung at planar scattering, averaged over $\bf{\theta}_{\textbf{k}}$ angles.}\\
\end{center}
\end{minipage}
\vspace{3mm}

\noindent Thus, without separation in $\bf{\theta}_{\textbf{k}}$, by
means of only keeping $\textbf{n}_{\textbf{q}_{\perp}}$ fixed,
polarization can not be obtained higher then 50\% (see Fig. 3). On
the other hand, we are going to show below, that with separation in
$\bf{\theta}_{\textbf{k}}$ performed, it is possible to achieve a
100\% polarization even at a spherically-symmetric scatterer.

\textbf{\textsc{3.2. Centrally-symmetric scatterer, azimuthal
integration over $\bf{q}_{\perp}$.}}

The averaging in (\ref{F}) over directions of $\bf{q}_{\perp}$ is
achieved through the substitution
$(n_{\textbf{q}_{\perp}})_{i}(n_{\textbf{q}_{\perp}})_{k}\rightarrow
\frac{1}{2}\delta_{ik}$. One gets
\begin{eqnarray*}
\left\langle \left| M_{rad} \right| ^{2} \right\rangle \rightarrow
\frac{8\textbf{q}_{\perp}^{2}}{m^{2}}\frac{(1-x_{\omega})^{2}}{(\gamma
^{-1}+\gamma \bf{\theta}_{\textbf{k}}^{2})^{2}}
\cdot&\left\{\frac{x_{\omega}^{2}}{2(1-x_{\omega})}+1-\frac{4\gamma^{-2}(\bf{\theta}_{\textbf{k}}\textbf{e}')^{2}}{(\gamma^{-2}+\bf{\theta}_{\textbf{k}})^{2}}
\right\},
\end{eqnarray*}
\begin{equation*}
P(x_{\omega},\gamma \left|\bf{\theta}_{\textbf{k}}
\right|)=\frac{2\gamma^{2}\bf{\theta}_{\textbf{k}}^{2}}{\frac{x_{\omega}^{2}}{2(1-x_{\omega})}(1+\gamma^{2}\bf{\theta}_{\textbf{k}}^{2})^{2}+1+\gamma^{4}\bf{\theta}_{\textbf{k}}^{4}}.
\end{equation*}
\\
\begin{minipage}{80mm}
\includegraphics[width=\textwidth]{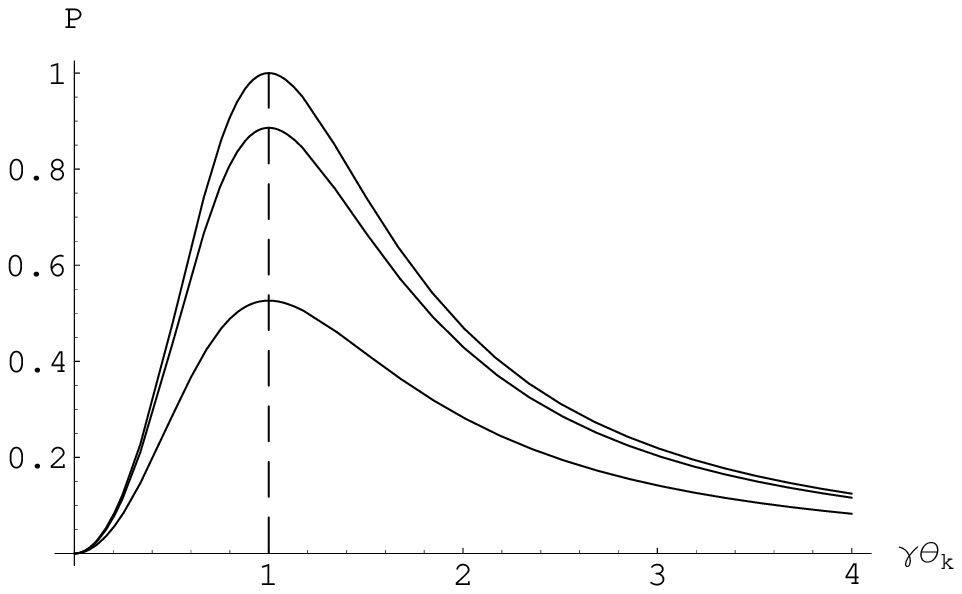}
\begin{center}
\emph{\textbf{Fig.4.}} {\it Polarization of bremsstrahlung on a
centrally-symmetric scatterer for $x_{\omega}=0$; $0.3$; $0.6$
\\(from
top to bottom).}\\
\end{center}
\end{minipage}
\vspace{3mm}

\noindent At $x_{\omega}\rightarrow 0$,
$\left|\bf{\theta}_{\textbf{k}} \right|=1/\gamma$  polarization
reaches 100\%, in spite of the angular averaging. That could hardly
be expected based on very general reasons only. (The corresponding
result was displayed in Fig. 4 of \cite{Ter-Mik} but with no
explanation supplied for the backing of such possibility). From our
Fig.1 it is apparent that 100\% magnitude of polarization may exist
because at $\left|\bf{\theta}_{\textbf{k}} \right|=1/\gamma$
polarization is oriented along a circle, centered at the origin of
the plane and invariant under rotations of
$\textbf{n}_{\textbf{q}_{\perp}}$, corresponding to angular
averaging.

Thus, observation at some radiation angles of polarization close to
100\% does not yet signal existence in the target of some special
collective fields, guiding electron motion. Such an effect is also
possible in an amorphous medium\footnote{As usual, at that in order
to be able to neglect multiple scattering effects as compared to
deflection by emitting radiation, one needs fulfillment of the
Landau-Pomeranchuk's type condition $L\ll e^{2}L_{rad}$ (see, e.g.,
\cite{BLP}), where $L$ is target thickness and $L_{rad}$ the
radiation length (centimeters to decimeters for solid targets).
Moreover, at $x_{\omega}$ rather small, photon emission angle which
is of main interest for us exceeds electron deflection angle by a
factor $\frac{1-x_{\omega}}{x_{\omega}}$, so the true condition may
be $L\ll
\left(\frac{1-x_{\omega}}{x_{\omega}}\right)^{2}e^{2}L_{rad}$.}.
\\
\begin{center}
\textbf{\textsc{4. SUMMARY}}
\end{center}
The present work has offered formulation of a method for evaluation
of differential cross-section of Compton scattering for polarized
initial and final photons, based on advanced reduction of the matrix
element. As was demonstrated, after the full reduction, the
calculation leading to Lorentz-invariant formula (\ref{KN1})
consumes only a few lines. To compare with, the derivation starting
with squaring of the initial matrix element or expression
(\ref{init}), occupies a few pages. The transition to differential
cross-section of bremsstrahlung in dipole approximation was based on
specification of covariant expressions in terms of vector components
in laboratory frame, which lifts the necessity to manually implement
the equivalent photon method and transit between different reference
frames - not very trivial task when dealing with polarization of the
emitted photon.

In application to bremsstrahlung, a previously overlooked feature
which seems to be worth emphasizing is that polarization as a
function of (small) radiation angles at fixed
$\textbf{n}_{\textbf{q}_{\perp}}$ orients itself along perfect
circles, including one centered at the origin. If averaging over
$\textbf{n}_{\textbf{q}_{\perp}}$ is performed, at the radius of the
latter circle $\left|\bf{\theta}_{\textbf{k}} \right|=1/\gamma$ only
polarizations with the same orientation add up. Moreover, at small
polarization values along the circle are close to 100\%, and so the
averaged polarization can be close to 100\%, too. Actually,
polarization stays in excess of 80\% in the interval
$0.8<\gamma\left|\mathbf{\theta}_{\textbf{k}} \right|<1.3$,
$x_{\omega}<0.3$.

The practical conclusions reached were as follows. If a beam of
energetic and polarized $\gamma$-quanta needs to be prepared, it is
beneficiary to use bremsstrahlung in an amorphous medium at
$\left|\bf{\theta}_{\textbf{k}} \right|\simeq 1/\gamma$ (not just
$\sim 1/\gamma)$. The polarization is orthogonal to the radiation
plane - as had long been established. The radiation recoil influence
is always depolarizing, but for $x_{\omega}<0.3$ rather weak. The
method is most convenient to use for $\omega \leq 10$ GeV.
\vspace{3mm}
\begin{center}

\end{center}
\end{document}